\documentclass[english]{article}
\usepackage[T1]{fontenc}
\usepackage[latin9]{inputenc}
\usepackage{geometry}
\usepackage{amsmath}
\usepackage{stmaryrd}

\makeatletter
\newcommand{\lyxaddress}[1]{
	\par {\raggedright #1
	\vspace{1.4em}
	\noindent\par}
}

\makeatother

\usepackage{babel}
\begin{document}
\title{\textbf{Analytical Approach of Matter Effect on (3+1) Neutrino Oscillation}}
\author{Vivek Kumar Nautiyal$\mathrm{^{1*}\,}\,$and~Bipin Singh Koranga$\mathrm{^{2}}$}
\maketitle

\lyxaddress{$\mathrm{^{1}}$Department of Physics, Babasaheb Bhimrao Ambedkar
University, Lucknow-226025, India}

\lyxaddress{$\mathrm{^{2}}$Department of Physics, Kirori Mal college (University
of Delhi), Delhi-110007, India}

$^{*}$Email - viveknautiyal01@gmail.com
\begin{abstract}
We study the analytic expression for four flavor neutrino oscillation
in the presence of matter. We calculate the time evolution operator
on flavor and mass basis. We find the matter dependent mass square
difference and neutrino transition probabilities for (3+1) four flavor
neutrino oscillation.
\end{abstract}

\section{Introduction}

At present there are four types of experiments for neutrino oscillation;
solar, atmospheric, accelerator and reactor neutrino experiments {[}1,
2, 3, 4, 5, 6, 7, 8{]}. The presence of sterile neutrino given by
LSND and MiniBoone results {[}7, 8{]}. LSND experiment gives solar
neutrino mass square difference $0.2<\Delta_{21}<100eV^{2}[9]$ and
MiniBooNe experiments gives solar mass square difference in the given
range $0.01<\Delta_{21}<1.0eV^{2}${[}10{]}. Both this experiments
gives new possibility of sterile neutrinos. The neutrino oscillation
parameter in vacuum can be modified by considering when neutrino interact
with matter. The well known MSW effects{[}11, 12{]}, which can be
explain by modified Hamiltonian. In four flavor neutrino oscilation,
the three active neutrinos interacts by neutral current interaction.
But the sterile neutrino has no any interaction.When neutrino passing
through in matter, the matter effect change the neutrino mass square
differences and mixing angles. Earlier many author work on three flavor
neutrino oscilation in matter{[}13, 14, 15, 18{]}. In this work,we
calculate the time evolution operator on four flavor neutrino oscillation.We
derive the expression for the matter dependent neutrino mass square
difference using Cayley--Hamilton theorem. The article outline is
as follows. In Section 2\textbf{,} four Flavor Neutrino Oscillation
in Vacuum. In Section 3, neutrino mass square difference and transition
probability in matter are driven for four flavor and the conclusion
is given in Section 4.

\section{Four Flavor Neutrino Oscillation in Vacuum}

In this section, we consider with four flavor framework (3+1) by assuming
the sterile neutrino of eV range and the mixing of this sterile neutrino
with three different neutrinos. By adding one sterile neutrinos {[}16{]},
there is an increment in mixing angles and CP violating phases in
the PMNS matrix $U_{4\vartimes4}$ which is given by {[}17{]},

\begin{equation}
U=R_{34}(\theta_{34},\delta_{34})R_{24}(\theta_{34})R_{14}(\theta_{14},\delta_{14})R_{23}(\theta_{23})R_{13}(\theta_{13},\delta_{13})R_{12}(\theta_{12}),
\end{equation}

where the matrices $R_{ij}$are rotations in ij space,

\[
R_{ij}(\theta_{ij},\delta)=\left(\begin{array}{cc}
c_{ij} & s_{ij}e^{-i\delta}\\
-s_{ij}e^{i\delta} & c_{ij}
\end{array}\right),
\]

where $s_{ij}=sin\theta_{ij}$,$\,c_{ij}=cos\theta_{ij}$.

Note that in four flavor there are three Dirac CP-violating phase
$\delta_{ij}.$ The explicit form of U is

\begin{equation}
U=\left(\begin{array}{cccc}
U_{e1} & U_{e2} & U_{e3} & U_{e4}\\
U_{\mu1} & U_{\mu2} & U_{\mu3} & U_{\mu4}\\
U_{\tau1} & U_{\tau2} & U_{\tau3} & U_{\tau4}\\
U_{s1} & U_{s2} & U_{s3} & U_{s4}
\end{array}\right),
\end{equation}

and

\[
\mathrm{U=\left(\begin{array}{cccc}
(c_{14}c_{13}c_{12}) & (c_{14}c_{13}s_{12}) & (c_{14}e^{-i\delta_{13}}s_{13}) & s_{14}e^{-i\delta_{13}}\\
\\
(-c_{24}c_{23}s_{12} & (c_{24}c_{23}c_{12} & (-c_{24}s_{23}c_{13} & (s_{24}c_{14})\\
-(c_{24}s_{23}s_{13}c_{12}e^{-i\delta_{13}} & -(c_{24}s_{23}s_{13}s_{12}e^{i\delta_{13}} & -s_{24}s_{14}s_{13}e^{-i(\delta_{13}-\delta_{14})})\\
-s_{24}s_{14}c_{13}c_{12}e^{-i\delta_{14}})) & -s_{24}s_{14}c_{13}s_{12}e^{i\delta_{14}}))\\
\\
(c_{34}s_{23}s_{12} & (-c_{34}s_{23}c_{12} & (c_{34}c_{23}c_{13} & (s_{34}c_{24}c_{14}e^{-i\delta_{34}})\\
-c_{34}c_{23}s_{13}c_{12}e^{i\delta_{13}} & -c_{34}c_{23}s_{13}s_{12}e^{i\delta_{13}} & -s_{34}s_{24}s_{23}c_{13}e^{-i\delta_{34}}\\
+s_{34}s_{24}c_{23}s_{12}e^{-i\delta_{34}} & -s_{34}s_{24}c_{23}c_{12}e^{-i\delta_{34}} & -s_{34}c_{24}s_{14}s_{13}e^{-i(\delta_{13}-\delta_{14}+\delta_{34})})\\
+s_{34}s_{24}s_{23}s_{13}c_{12}e^{-i(\delta_{34}-\delta_{13})} & +s_{34}s_{24}s_{23}s_{13}s_{12}e^{-i(\delta_{34}-\delta_{13})}\\
-s_{34}c_{24}s_{14}c_{13}c_{12}e^{-i(\delta_{34}-\delta_{14})}) & -s_{34}c_{24}s_{14}c_{13}s_{12}e^{-i(\delta_{34}-\delta_{14})})\\
\\
(-s_{34}s_{23}s_{12}e^{i\delta_{34}} & (s_{34}s_{23}c_{12}e^{i\delta_{34}} & (-s_{34}c_{23}c_{13}e^{i\delta_{34}} & (c_{34}c_{24}c_{14})\\
+s_{34}c_{23}s_{13}c_{12}e^{i(\delta_{34}+\delta_{13})} & +s_{34}c_{23}s_{13}s_{12}e^{i(\delta_{34}+\delta_{13})} & -c_{34}s_{24}s_{23}c_{13}\\
+c_{34}s_{24}c_{23}s_{12} & -c_{34}s_{24}c_{23}c_{12} & -c_{34}c_{24}s_{14}s_{13}e^{-i(\delta_{13}-\delta_{14})})\\
+c_{34}s_{24}s_{23}s_{12}c_{12}e^{i\delta_{13}} & +c_{34}s_{24}s_{23}s_{13}s_{12}e^{i\delta_{13}}\\
-c_{34}c_{24}s_{14}c_{13}c_{12}e^{i\delta_{14}}) & -c_{34}c_{24}s_{14}c_{13}s_{12}e^{i\delta_{14}})
\end{array}\right)}
\]

In the presence of one sterile neutrino {[}19{]}, the four flavor
neutrino oscillation probability in vacuum is

\begin{equation}
P_{\nu_{\alpha}\rightarrow\nu_{\beta}}=\delta_{\alpha\beta}-4\sum_{i<j}^{4}Re(U_{\alpha i}U_{\beta j}U_{\alpha j}^{*}U_{\beta i}^{*})sin^{2}\bar{\Delta}_{ij}+2\sum_{i<j}^{4}Im(U_{\alpha i}U_{\beta j}U_{\alpha j}^{*}U_{\beta i}^{*})sin\bar{2\Delta}_{ij},\,\,\,\,\,\,\,\alpha,\beta=e,\mu,\tau,s
\end{equation}

where $\bar{\Delta}_{ij}=\Delta_{ij}L/2E$ , baseline length of particular
experiment is L.

\section{Neutrino Mass Square Difference in Matter for Four Flavor}

If we assume the mass eigenstate basis, the Hamiltonian $H_{vacuum}$
in the propagation of neutrinos in vacuum is given by

\begin{equation}
H_{vaccum}=\left(\begin{array}{cccc}
E_{1} & 0 & 0 & 0\\
0 & E_{2} & 0 & 0\\
0 & 0 & E_{3} & 0\\
0 & 0 & 0 & E_{4}
\end{array}\right),
\end{equation}

where $E_{k}(k=1,2,3,4)$ are the energies of the neutrino mass eigenstates
k with mass $m_{k};$

\begin{equation}
E_{k}=\sqrt{m_{k}^{2}+p_{k}^{2}}\approx p_{k}+\frac{m_{k}^{2}}{2p}\approx p+\frac{m_{k}^{2}}{2E}
\end{equation}

we consider the momentum p is the same for all mass eigenstates. When
neutrino interact with matter by weak interaction (charged and neutral
current). Sterile neutrino itself not take any participitation of
weak interaction. The effective Hamiltonian for four flavor neutrino
mixing is {[}12{]}

\begin{equation}
H_{vaccum}=\frac{1}{2E}\left[\left(\begin{array}{cccc}
m_{1}^{2} & 0 & 0 & 0\\
0 & m_{2}^{2} & 0 & 0\\
0 & 0 & m_{3}^{2} & 0\\
0 & 0 & 0 & m_{4}^{2}
\end{array}\right)\right],
\end{equation}

When neutrinos propagate in matter, there is an additional term coming
from the presence of electrons in the matter {[}5{]}. This term is
diagonal in the flavor state basis and is given by

\begin{equation}
A_{f}=\frac{1}{2E}\left(\begin{array}{cccc}
A & 0 & 0 & 0\\
0 & 0 & 0 & 0\\
0 & 0 & 0 & 0\\
0 & 0 & 0 & A^{'}
\end{array}\right)
\end{equation}

where,

\[
A(eV^{2})=2\sqrt{2}G_{F}N_{e}E_{\nu},
\]

\[
A^{'}(eV^{2})=-\sqrt{2}G_{F}N_{n}E_{\nu},
\]

and $N_{e}$ and $N_{n}$ is the density of electron and neutron {[}17{]}.
Since, we can switch from flavor state to mass eigen state via transformation.
Thus, the matter induce Hamiltonian {[}17, 18{]} is $\,H_{m}=U^{T}A_{f}U$,
where U is the PMNS matrix. Now the total Hamiltonian is given by,
\begin{equation}
H_{m}=U^{T}A_{f}U
\end{equation}
\begin{equation}
H=H_{vaccum}+H_{m}
\end{equation}
\begin{equation}
\frac{1}{2E}\left[\widetilde{U}^{T}\left(\begin{array}{cccc}
\widetilde{m}_{1}^{2} & 0 & 0 & 0\\
0 & \widetilde{m}_{2}^{2} & 0 & 0\\
0 & 0 & \widetilde{m}_{3}^{2} & 0\\
0 & 0 & 0 & \widetilde{m}_{4}^{2}
\end{array}\right)\widetilde{U}\right]=H=\frac{1}{2E}\left[\left(\begin{array}{cccc}
m_{1}^{2} & 0 & 0 & 0\\
0 & m_{2}^{2} & 0 & 0\\
0 & 0 & m_{3}^{2} & 0\\
0 & 0 & 0 & m_{4}^{2}
\end{array}\right)+U^{T}\left(\begin{array}{cccc}
A & 0 & 0 & 0\\
0 & 0 & 0 & 0\\
0 & 0 & 0 & 0\\
0 & 0 & 0 & A^{'}
\end{array}\right)U\right]
\end{equation}

\begin{equation}
\left[\widetilde{U}^{T}\left(\begin{array}{cccc}
\widehat{m}_{11}^{2} & 0 & 0 & 0\\
0 & \widehat{m}_{21}^{2} & 0 & 0\\
0 & 0 & \widehat{m}_{31}^{2} & 0\\
0 & 0 & 0 & \widehat{m}_{41}^{2}
\end{array}\right)\widetilde{U}\right]=H=\left[\left(\begin{array}{cccc}
0 & 0 & 0 & 0\\
0 & \Delta_{21} & 0 & 0\\
0 & 0 & \Delta_{31} & 0\\
0 & 0 & 0 & \Delta_{41}
\end{array}\right)+U^{T}\left(\begin{array}{cccc}
A & 0 & 0 & 0\\
0 & 0 & 0 & 0\\
0 & 0 & 0 & 0\\
0 & 0 & 0 & A^{'}
\end{array}\right)U\right]
\end{equation}
where,
\[
\widehat{m}_{ij}^{2}=\widetilde{m}_{i}^{2}-m_{j}^{2},\,\,\,\,\,\,\,\,and\,\,\,\,\,\,\,\,\,\,\,\Delta_{ij}=m_{i}^{2}-m_{j}^{2}
\]

The explicit form of Hamiltonian are,

\begin{equation}
H_{m}=\left(\begin{array}{cccc}
AU_{e1}^{2}+A^{'}U_{s1}^{2} & AU_{e1}U_{e2}+A^{'}U_{s1}U_{s2} & AU_{e1}U_{e3}+A^{'}U_{s1}U_{s3} & AU_{e1}U_{e4}+A^{'}U_{s1}U_{s4}\\
AU_{e1}U_{e2}+A^{'}U_{s1}U_{s2} & AU_{e2}^{2}+A^{'}U_{s2}^{2} & AU_{e2}U_{e3}+A^{'}U_{s2}U_{s3} & AU_{e2}U_{e4}+A^{'}U_{s2}U_{s4}\\
AU_{e1}U_{e3}+A^{'}U_{s1}U_{s3} & AU_{e2}U_{e3}+A^{'}U_{s2}U_{s3} & AU_{e3}^{2}+A^{'}U_{s3}^{2} & AU_{e3}U_{e4}+A^{'}U_{s3}U_{s4}\\
AU_{e1}U_{e4}+A^{'}U_{s1}U_{s4} & AU_{e2}U_{e4}+A^{'}U_{s2}U_{s4} & AU_{e3}U_{e4}+A^{'}U_{s3}U_{s4} & AU_{e4}^{2}+A^{'}U_{s4}^{2}
\end{array}\right)
\end{equation}

\begin{equation}
H=\left(\begin{array}{cccc}
AU_{e1}^{2}+A^{'}U_{s1}^{2} & AU_{e1}U_{e2}+A^{'}U_{s1}U_{s2} & AU_{e1}U_{e3}+A^{'}U_{s1}U_{s3} & AU_{e1}U_{e4}+A^{'}U_{s1}U_{s4}\\
AU_{e1}U_{e2}+A^{'}U_{s1}U_{s2} & \Delta_{21}+AU_{e2}^{2}+A^{'}U_{s2}^{2} & AU_{e2}U_{e3}+A^{'}U_{s2}U_{s3} & AU_{e2}U_{e4}+A^{'}U_{s2}U_{s4}\\
AU_{e1}U_{e3}+A^{'}U_{s1}U_{s3} & AU_{e2}U_{e3}+A^{'}U_{s2}U_{s3} & \Delta_{31}+AU_{e3}^{2}+A^{'}U_{s3}^{2} & AU_{e3}U_{e4}+A^{'}U_{s3}U_{s4}\\
AU_{e1}U_{e4}+A^{'}U_{s1}U_{s4} & AU_{e2}U_{e4}+A^{'}U_{s2}U_{s4} & AU_{e3}U_{e4}+A^{'}U_{s3}U_{s4} & \Delta_{41}+AU_{e4}^{2}+A^{'}U_{s4}^{2}
\end{array}\right)
\end{equation}
The propogation of neutrinos or change of flavor in oscillation at
some distance L is govern by the time evolution of a wavefunction
$\,\varPhi_{m}(t)\,$, which is basically a plane wave solution of
schrodinger equation, given by,
\begin{equation}
\varPhi_{m}(x,t)=e^{-iHt}\varphi_{m}(x)
\end{equation}
where,$\,\varphi_{m}(t)\,,H_{m}\,$ are the wavefunction and Hamiltonian
in mass eigenstate basis. At $\,t=L\,,$we have,
\begin{equation}
\varPhi_{m}(x,L)=e^{-iHL}\varphi_{m}(x)
\end{equation}
We can simply switch to flavor eigen basis by applying the unitary
operator. Then eq. (15) becomes,
\[
\varPhi_{m}(x,t)=Ue^{-iHL}\varphi_{m}(x)
\]
\[
\varPhi_{f}(x,t)=Ue^{-iHL}U^{T}U\varphi_{m}(x)
\]
\begin{equation}
\varPhi_{f}(x,t)=e^{-iH_{f}L}\varphi_{f}(x)
\end{equation}

Now we are finding the exponential terms in mass eigenstate and flavor
eigenstate basis,$\,e^{-iHL}\,\,and\,\,e^{-iH_{f}L},$~respectively.
We can write any $\,4\times4\,$matrix in term of traceless matrix,
i.e.

\begin{equation}
A=A_{0}+\frac{1}{4}(Trace\,A)I
\end{equation}

where,$\,A_{0}\,$is a matrix with zero trace. Then,

\[
e^{-iHL}=\varOmega e^{-iLT}=e^{-iHL-\frac{1}{4}(Trace\,H)IL}
\]

where,
\[
\varOmega=e^{-iL\frac{1}{4}(Trace\,H)}\,\,\,\,\,\,\,\,and\,\,\,\,\,\,\,\,T=H-\frac{1}{4}(Trace\,H)I
\]

\begin{equation}
e^{-iLT}=\sum_{n=1}^{4}\frac{(-iLT)^{n}}{n!}=-iLT-\frac{L^{2}T^{2}}{2!}+i\frac{L^{3}T^{3}}{3!}+\frac{L^{4}T^{4}}{4!}+........
\end{equation}
The characterstic equation of matrix $\,H\,$ is given by,
\begin{equation}
det(-iHL-\lambda I)=0
\end{equation}
\begin{equation}
\lambda^{n}+a_{n-1}\lambda^{n-1}+.......+a_{1}\lambda+a_{0}I=0
\end{equation}
Using Cayley-Hamilton's theorem $\,\lambda\,$is change by matrix
$\,H$, i.e
\[
H^{n}+a_{n-1}H^{n-1}+.......+a_{1}H+a_{0}I=0
\]
\[
H^{n}=-a_{n-1}H^{n-1}-.......-a_{1}H-a_{0}I
\]
\begin{equation}
H^{n}=c_{n-1}H^{n-1}+.......+c_{1}H+c_{0}I
\end{equation}
Using eq. (18) and eq. (21), we get

\[
e^{-iLT}=c_{0}I+c_{1}(-iLT)+c_{2}(-iLT)^{2}+c_{3}(-iLT)^{3}
\]
\begin{equation}
e^{-iLT}=c_{0}I-ic_{1}(LT)-c_{2}L^{2}T^{2}+ic_{3}L^{3}T^{3}
\end{equation}

\begin{equation}
e^{-iHL}=\varOmega\left(c_{0}I-ic_{1}(LT)-c_{2}L^{2}T^{2}+ic_{3}L^{3}T^{3}\right)
\end{equation}
The Trace of Matrix H is,
\[
Trace\,H=\Delta_{21}+\Delta_{31}+\Delta_{41}+A+A^{'}
\]
and the matrix T is given by,
\begin{equation}
T=\left(\begin{array}{cccc}
T_{11} & T_{12} & T_{13} & T_{14}\\
T_{21} & T_{22} & T_{23} & T_{24}\\
T_{31} & T_{32} & T_{33} & T_{34}\\
T_{41} & T_{42} & T_{43} & T_{44}
\end{array}\right)
\end{equation}
With,
\begin{equation}
T_{11}=AU_{e1}^{2}+A^{'}U_{s1}^{2}-\frac{1}{4}(\Delta_{21}+\Delta_{31}+\Delta_{41})-\frac{1}{4}(A+A^{'})
\end{equation}

\begin{equation}
T_{22}=AU_{e2}^{2}+A^{'}U_{s2}^{2}-\frac{1}{4}(-\Delta_{21}+\Delta_{32}+\Delta_{42})-\frac{1}{4}(A+A^{'})
\end{equation}
\begin{equation}
T_{33}=AU_{e3}^{2}+A^{'}U_{s3}^{2}-\frac{1}{4}(-\Delta_{32}-\Delta_{31}+\Delta_{43})-\frac{1}{4}(A+A^{'})
\end{equation}
\begin{equation}
T_{44}=AU_{e4}^{2}+A^{'}U_{s4}^{2}-\frac{1}{4}(-\Delta_{42}-\Delta_{43}-\Delta_{41})-\frac{1}{4}(A+A^{'})
\end{equation}

And
\begin{equation}
\begin{array}{c}
T_{12}=T_{21}=AU_{e1}U_{e2}+A^{'}U_{s1}U_{s2}\\
T_{13}=T_{31}=AU_{e1}U_{e3}+A^{'}U_{s1}U_{s3}\\
T_{14}=T_{41}=AU_{e1}U_{e4}+A^{'}U_{s1}U_{s4}\\
T_{23}=T_{32}=AU_{e2}U_{e3}+A^{'}U_{s2}U_{s3}\\
T_{24}=T_{42}=AU_{e2}U_{e4}+A^{'}U_{s2}U_{s4}\\
T_{34}=T_{43}=AU_{e3}U_{e4}+A^{'}U_{s3}U_{s4}
\end{array}
\end{equation}

We can see from eq. (25-29) that the matrix $\,T\,$is Traceless symmetrix
matrix.i.e
\begin{equation}
\begin{array}{ccccc}
\sum_{i=1}^{4}T_{ii}=0 &  & ; &  & T_{ij}=T_{ji}\end{array}
\end{equation}

Characterstic equation of matrix T is
\begin{equation}
a_{4}\lambda^{4}+a_{3}\lambda^{3}+a_{2}\lambda^{2}+a_{1}\lambda+a_{0}=0
\end{equation}

where the coefficients $a_{4},\,a_{3},\,a_{2},\,a_{1}\,and\,a_{0}\,$are
given in Appendix A.

\[
\begin{array}{ccccc}
P=\frac{3a_{3}^{2}-8a_{2}a_{4}}{12a_{4}^{2}} &  & , &  & X=-a_{3}^{3}+4a_{3}a_{2}a_{4}-8a_{1}a_{4}^{2},\end{array}
\]

\[
\begin{array}{ccccc}
L=a_{2}^{2}-3a_{3}a_{1}+12a_{0}a_{4} &  & , &  & M=2a_{2}^{3}-9a_{3}a_{1}a_{2}+27a_{3}^{2}a_{0}+27a_{1}^{2}a_{4}-72a_{2}a_{0}a_{4},\end{array}
\]

\[
Q=\sqrt{P+\frac{1}{3a_{4}}\left[\left(\frac{M}{2}+\sqrt{\left(\frac{M}{2}\right)^{2}-L^{3}}\right)^{1/3}+L\left(\frac{M}{2}+\sqrt{\left(\frac{M}{2}\right)^{2}-L^{3}}\right)^{-1/3}\right]}
\]

The roots of eq.(31), gives matter dependent mass square $m_{m1}^{2},m_{m2}^{2},m_{m3}^{2},m_{m4}^{2}$

\[
\lambda_{1}=m_{m1}^{2}=-\frac{a_{3}}{4}-\frac{Q}{2}-\frac{1}{2}\sqrt{3P-Q^{2}-\frac{X}{4Q}},
\]

\[
\lambda_{2}=m_{m2}^{2}=-\frac{a_{3}}{4}-\frac{Q}{2}+\frac{1}{2}\sqrt{3P-Q^{2}-\frac{X}{4Q}},
\]
\[
\lambda_{3}=m_{m3}^{2}=-\frac{a_{3}}{4}+\frac{Q}{2}-\frac{1}{2}\sqrt{3P-Q^{2}+\frac{X}{4Q}},
\]

\[
\lambda_{4}=m_{m4}^{2}=-\frac{a_{3}}{4}+\frac{Q}{2}+\frac{1}{2}\sqrt{3P-Q^{2}+\frac{X}{4Q}}.
\]

Using matter dependent mass square $m_{m1}^{2},m_{m2}^{2},m_{m3}^{2},m_{m4}^{2}$,we
can write matter dependent mass square difference for four flavor
neutrino oscillation

\begin{equation}
\Delta_{21}^{m}=m_{m2}^{2}-m_{m1}^{2}=\sqrt{3P-Q^{2}-\frac{X}{4Q}},
\end{equation}

\begin{equation}
\Delta_{31}^{m}=m_{m3}^{2}-m_{m1}^{2}=Q+\frac{1}{2}\left(\sqrt{3P-Q^{2}-\frac{X}{4Q}}-\sqrt{3P-Q^{2}+\frac{X}{4Q}}\right),
\end{equation}

\begin{equation}
\Delta_{41}^{m}=m_{m4}^{2}-m_{m1}^{2}=Q+\frac{1}{2}\left(\sqrt{3P-Q^{2}-\frac{X}{4Q}}+\sqrt{3P-Q^{2}+\frac{X}{4Q}}\right).
\end{equation}
Now, from eq.(22) we have
\[
e^{-iL\lambda_{1}}=c_{0}I-ic_{1}L\lambda_{1}-c_{2}L^{2}\lambda_{1}^{2}+ic_{3}L^{3}\lambda_{1}^{3}
\]
\[
e^{-iL\lambda_{2}}=c_{0}I-ic_{1}L\lambda_{2}-c_{2}L^{2}\lambda_{2}^{2}+ic_{3}L^{3}\lambda_{2}^{3}
\]
\[
e^{-iL\lambda_{3}}=c_{0}I-ic_{1}L\lambda_{3}-c_{2}L^{2}\lambda_{3}^{2}+ic_{3}L^{3}\lambda_{3}^{3}
\]
\[
e^{-iL\lambda_{4}}=c_{0}I-ic_{1}L\lambda_{4}-c_{2}L^{2}\lambda_{4}^{2}+ic_{3}L^{3}\lambda_{4}^{3}
\]

\begin{equation}
\left(\begin{array}{c}
e^{-iL\lambda_{1}}\\
e^{-iL\lambda_{2}}\\
e^{-iL\lambda_{3}}\\
e^{-iL\lambda_{4}}
\end{array}\right)=\left[\begin{array}{cccc}
1 & -iL\lambda_{1} & -L^{2}\lambda_{1}^{2} & iL^{3}\lambda_{1}^{3}\\
1 & -iL\lambda_{2} & -L^{2}\lambda_{2}^{2} & iL^{3}\lambda_{2}^{3}\\
1 & -iL\lambda_{3} & -L^{2}\lambda_{3}^{2} & iL^{3}\lambda_{3}^{3}\\
1 & -iL\lambda_{4} & -L^{2}\lambda_{4}^{2} & iL^{3}\lambda_{4}^{3}
\end{array}\right]\left(\begin{array}{c}
c_{0}\\
c_{1}\\
c_{2}\\
c_{3}
\end{array}\right)
\end{equation}
\begin{equation}
\left(\begin{array}{c}
c_{0}\\
c_{1}\\
c_{2}\\
c_{3}
\end{array}\right)=\left[\begin{array}{cccc}
1 & -iL\lambda_{1} & -L^{2}\lambda_{1}^{2} & iL^{3}\lambda_{1}^{3}\\
1 & -iL\lambda_{2} & -L^{2}\lambda_{2}^{2} & iL^{3}\lambda_{2}^{3}\\
1 & -iL\lambda_{3} & -L^{2}\lambda_{3}^{2} & iL^{3}\lambda_{3}^{3}\\
1 & -iL\lambda_{4} & -L^{2}\lambda_{4}^{2} & iL^{3}\lambda_{4}^{3}
\end{array}\right]^{-1}\left(\begin{array}{c}
e^{-iL\lambda_{1}}\\
e^{-iL\lambda_{2}}\\
e^{-iL\lambda_{3}}\\
e^{-iL\lambda_{4}}
\end{array}\right)
\end{equation}
\textbf{
\begin{equation}
\boldsymbol{c=\mathcal{L}.e}
\end{equation}
}

Where,
\begin{equation}
\begin{array}{c}
\mathcal{L}_{11}=\frac{1}{D}(L^{6}(\lambda_{4}(\lambda_{2}^{3}\lambda_{3}^{2}-\lambda_{2}^{2}\lambda_{3}^{3})+\lambda_{3}(\lambda_{2}^{2}\lambda_{4}^{3}-\lambda_{2}^{3}\lambda_{4}^{2})+\lambda_{2}(\lambda_{3}^{3}\lambda_{4}^{2}-\lambda_{3}^{2}\lambda_{4}^{3})))\\
\mathcal{L}_{12}=\frac{1}{D}(L^{6}(\lambda_{4}(\lambda_{3}^{3}\lambda_{1}^{2}-\lambda_{3}^{2}\lambda_{1}^{3})+\lambda_{3}(\lambda_{4}^{2}\lambda_{1}^{3}-\lambda_{4}^{3}\lambda_{1}^{2})+\lambda_{1}(\lambda_{4}^{3}\lambda_{3}^{2}-\lambda_{4}^{2}\lambda_{3}^{3})))\\
\mathcal{L}_{13}=\frac{1}{D}(L^{6}(\lambda_{4}(\lambda_{2}^{2}\lambda_{1}^{3}-\lambda_{1}^{2}\lambda_{2}^{3})+\lambda_{2}(\lambda_{1}^{2}\lambda_{4}^{3}-\lambda_{1}^{3}\lambda_{4}^{2})+\lambda_{1}(\lambda_{2}^{3}\lambda_{4}^{2}-\lambda_{2}^{2}\lambda_{4}^{3})))\\
\mathcal{L}_{14}=\frac{1}{D}(L^{6}(\lambda_{3}(\lambda_{1}^{3}\lambda_{2}^{2}-\lambda_{1}^{2}\lambda_{2}^{3})+\lambda_{2}(\lambda_{3}^{2}\lambda_{1}^{3}-\lambda_{3}^{3}\lambda_{1}^{2})+\lambda_{1}(\lambda_{3}^{3}\lambda_{2}^{2}-\lambda_{3}^{2}\lambda_{2}^{3})))\\
\mathcal{L}_{21}=\frac{1}{D}(iL^{5}((\lambda_{3}^{3}\lambda_{2}^{2}-\lambda_{3}^{2}\lambda_{2}^{3})+(\lambda_{4}^{2}\lambda_{2}^{3}-\lambda_{4}^{3}\lambda_{2}^{2})+(\lambda_{4}^{3}\lambda_{3}^{2}-\lambda_{4}^{2}\lambda_{3}^{3})))\\
\mathcal{L}_{22}=\frac{1}{D}(iL^{5}((\lambda_{1}^{3}\lambda_{3}^{2}-\lambda_{1}^{2}\lambda_{3}^{3})+(\lambda_{1}^{2}\lambda_{4}^{3}-\lambda_{1}^{3}\lambda_{4}^{2})+(\lambda_{3}^{3}\lambda_{4}^{2}-\lambda_{3}^{2}\lambda_{4}^{3})))\\
\mathcal{L}_{23}=\frac{1}{D}(iL^{5}((\lambda_{2}^{3}\lambda_{1}^{2}-\lambda_{2}^{2}\lambda_{1}^{3})+(\lambda_{4}^{2}\lambda_{1}^{3}-\lambda_{4}^{3}\lambda_{1}^{2})+(\lambda_{4}^{3}\lambda_{2}^{2}-\lambda_{4}^{2}\lambda_{2}^{3})))\\
\mathcal{L}_{24}=\frac{1}{D}(iL^{5}((\lambda_{1}^{3}\lambda_{2}^{2}-\lambda_{1}^{2}\lambda_{2}^{3})+(\lambda_{1}^{2}\lambda_{3}^{3}-\lambda_{1}^{3}\lambda_{3}^{2})+(\lambda_{2}^{3}\lambda_{3}^{2}-\lambda_{2}^{2}\lambda_{3}^{3})))\\
\mathcal{L}_{31}=\frac{1}{D}(L^{4}((\lambda_{3}^{3}\lambda_{2}-\lambda_{2}^{3}\lambda_{3})+(\lambda_{2}^{3}\lambda_{4}-\lambda_{4}^{3}\lambda_{2})+(\lambda_{4}^{3}\lambda_{3}-\lambda_{3}^{3}\lambda_{4})))\\
\mathcal{L}_{32}=\frac{1}{D}(L^{4}((\lambda_{1}^{3}\lambda_{3}-\lambda_{3}^{3}\lambda_{1})+(\lambda_{4}^{3}\lambda_{1}-\lambda_{1}^{3}\lambda_{4})+(\lambda_{3}^{3}\lambda_{4}-\lambda_{4}^{3}\lambda_{3})))\\
\mathcal{L}_{33}=\frac{1}{D}(L^{4}((\lambda_{2}^{3}\lambda_{1}-\lambda_{1}^{3}\lambda_{2})+(\lambda_{1}^{3}\lambda_{4}-\lambda_{4}^{3}\lambda_{1})+(\lambda_{4}^{3}\lambda_{2}-\lambda_{2}^{3}\lambda_{4})))\\
\mathcal{L}_{34}=\frac{1}{D}(L^{4}((\lambda_{1}^{3}\lambda_{2}-\lambda_{2}^{3}\lambda_{1})+(\lambda_{3}^{3}\lambda_{1}-\lambda_{1}^{3}\lambda_{3})+(\lambda_{2}^{3}\lambda_{3}-\lambda_{3}^{3}\lambda_{2})))\\
\mathcal{L}_{41}=\frac{1}{D}(iL^{3}((\lambda_{2}^{2}\lambda_{3}-\lambda_{3}^{2}\lambda_{2})+(\lambda_{4}^{2}\lambda_{2}-\lambda_{2}^{2}\lambda_{4})+(\lambda_{3}^{2}\lambda_{4}-\lambda_{4}^{2}\lambda_{3})))\\
\mathcal{L}_{42}=\frac{1}{D}(iL^{3}((\lambda_{3}^{2}\lambda_{1}-\lambda_{1}^{2}\lambda_{3})+(\lambda_{1}^{2}\lambda_{4}-\lambda_{4}^{2}\lambda_{1})+(\lambda_{4}^{2}\lambda_{3}-\lambda_{3}^{2}\lambda_{4})))\\
\mathcal{L}_{43}=\frac{1}{D}(iL^{3}((\lambda_{1}^{2}\lambda_{2}-\lambda_{2}^{2}\lambda_{1})+(\lambda_{1}^{2}\lambda_{4}-\lambda_{4}^{2}\lambda_{1})+(\lambda_{2}^{2}\lambda_{4}-\lambda_{4}^{2}\lambda_{2})))\\
\mathcal{L}_{44}=\frac{1}{D}(iL^{3}((\lambda_{2}^{2}\lambda_{1}-\lambda_{1}^{2}\lambda_{2})+(\lambda_{1}^{2}\lambda_{3}-\lambda_{3}^{2}\lambda_{1})+(\lambda_{3}^{2}\lambda_{2}-\lambda_{2}^{2}\lambda_{3})))
\end{array}
\end{equation}
with,

\begin{equation}
\begin{array}{c}
D=(L^{6}(\lambda_{4}(\lambda_{1}^{3}\lambda_{2}^{2}+\lambda_{2}^{3}\lambda_{3}^{2}+\lambda_{3}^{3}\lambda_{1}^{2}-\lambda_{1}^{2}\lambda_{2}^{3}-\lambda_{3}^{2}\lambda_{1}^{3}-\lambda_{2}^{2}\lambda_{3}^{3})\\
+\lambda_{3}(\lambda_{2}^{3}\lambda_{1}^{2}+\lambda_{1}^{3}\lambda_{4}^{2}+\lambda_{4}^{3}\lambda_{2}^{2}-\lambda_{2}^{2}\lambda_{1}^{3}-\lambda_{4}^{2}\lambda_{2}^{3}-\lambda_{1}^{2}\lambda_{4}^{3})\\
+\lambda_{2}(\lambda_{1}^{3}\lambda_{3}^{2}+\lambda_{3}^{3}\lambda_{4}^{2}+\lambda_{4}^{3}\lambda_{2}^{2}-\lambda_{1}^{2}\lambda_{4}^{3}-\lambda_{4}^{2}\lambda_{3}^{3}-\lambda_{4}^{2}\lambda_{1}^{3})\\
+\lambda_{1}(\lambda_{3}^{3}\lambda_{2}^{2}+\lambda_{2}^{3}\lambda_{4}^{2}+\lambda_{4}^{3}\lambda_{3}^{2}-\lambda_{3}^{2}\lambda_{2}^{3}-\lambda_{4}^{2}\lambda_{2}^{3}-\lambda_{4}^{2}\lambda_{3}^{3})))
\end{array}
\end{equation}
Then coefficients are,
\begin{equation}
c_{0}=\mathcal{L}_{11}e^{-iL\lambda_{1}}+\mathcal{L}_{12}e^{-iL\lambda_{2}}+\mathcal{L}_{13}e^{-iL\lambda_{3}}+\mathcal{L}_{14}e^{-iL\lambda_{4}}
\end{equation}
\begin{equation}
c_{1}=\mathcal{L}_{21}e^{-iL\lambda_{1}}+\mathcal{L}_{22}e^{-iL\lambda_{2}}+\mathcal{L}_{23}e^{-iL\lambda_{3}}+\mathcal{L}_{24}e^{-iL\lambda_{4}}
\end{equation}
\begin{equation}
c_{2}=\mathcal{L}_{31}e^{-iL\lambda_{1}}+\mathcal{L}_{32}e^{-iL\lambda_{2}}+\mathcal{L}_{33}e^{-iL\lambda_{3}}+\mathcal{L}_{34}e^{-iL\lambda_{4}}
\end{equation}
\begin{equation}
c_{3}=\mathcal{L}_{41}e^{-iL\lambda_{1}}+\mathcal{L}_{42}e^{-iL\lambda_{2}}+\mathcal{L}_{43}e^{-iL\lambda_{3}}+\mathcal{L}_{44}e^{-iL\lambda_{4}}
\end{equation}
Using eq. (40-43) , eq.(23) becomes
\begin{equation}
\begin{array}{c}
e^{-iHL}=\varOmega(e^{-iL\lambda_{1}}(\mathcal{L}_{11}I-i\mathcal{L}_{21}LT-\mathcal{L}_{31}L^{2}T^{2}+i\mathcal{L}_{41}L^{3}T^{3})\\
+e^{-iL\lambda_{2}}(\mathcal{L}_{12}I-i\mathcal{L}_{22}LT-\mathcal{L}_{32}L^{2}T^{2}+i\mathcal{L}_{42}L^{3}T^{3})\\
+e^{-iL\lambda_{3}}(\mathcal{L}_{13}I-i\mathcal{L}_{23}LT-\mathcal{L}_{33}L^{2}T^{2}+i\mathcal{L}_{43}L^{3}T^{3})\\
+e^{-iL\lambda_{4}}(\mathcal{L}_{14}I-i\mathcal{L}_{24}LT-\mathcal{L}_{34}L^{2}T^{2}+i\mathcal{L}_{44}L^{3}T^{3}))
\end{array}
\end{equation}
\begin{equation}
\begin{array}{c}
e^{-iHL}=\varOmega(e^{-iL\lambda_{1}}(\mathcal{L}_{11}I+\mathcal{L}_{21}T-\mathcal{L}_{31}T^{2}-\mathcal{L}_{41}T^{3})\\
+e^{-iL\lambda_{2}}(\mathcal{L}_{21}I+\mathcal{L}_{22}T-\mathcal{L}_{32}T^{2}-\mathcal{L}_{42}T^{3})\\
+e^{-iL\lambda_{3}}(\mathcal{L}_{13}I+\mathcal{L}_{23}T-\mathcal{L}_{33}T^{2}-\mathcal{L}_{43}T^{3})\\
+e^{-iL\lambda_{4}}(\mathcal{L}_{14}I+\mathcal{L}_{24}T-\mathcal{L}_{34}T^{2}-\mathcal{L}_{44}T^{3}))
\end{array}
\end{equation}
The above equation is the evolution operator in mass eigenstate basis.
Thus the evolution operator for the neutrinos in the flavor basis
is given by,
\begin{equation}
e^{-iH_{f}L}=Ue^{-iHL}U^{T}
\end{equation}
\begin{equation}
\begin{array}{c}
e^{-iH_{f}L}=\varOmega(e^{-iL\lambda_{1}}(\mathcal{L}_{11}I+\mathcal{L}_{21}\widehat{T}-\mathcal{L}_{31}\widehat{T}^{2}-\mathcal{L}_{41}\widehat{T}^{3})\\
+e^{-iL\lambda_{2}}(\mathcal{L}_{12}I+\mathcal{L}_{22}\widehat{T}-\mathcal{L}_{32}\widehat{T}^{2}-\mathcal{L}_{42}\widehat{T}^{3})\\
+e^{-iL\lambda_{3}}(\mathcal{L}_{13}I+\mathcal{L}_{23}\widehat{T}-\mathcal{L}_{33}\widehat{T}^{2}-\mathcal{L}_{43}\widehat{T}^{3})\\
+e^{-iL\lambda_{4}}(\mathcal{L}_{14}I+\mathcal{L}_{24}\widehat{T}-\mathcal{L}_{34}\widehat{T}^{2}-\mathcal{L}_{44}\widehat{T}^{3}))
\end{array}
\end{equation}
\begin{equation}
e^{-iH_{f}L}=\varOmega\sum_{j=1}^{4}(e^{-iL\lambda_{j}}(\mathcal{L}_{1j}I+\mathcal{L}_{2j}\widehat{T}-\mathcal{L}_{3j}\widehat{T}^{2}-\mathcal{L}_{4j}\widehat{T}^{3}))
\end{equation}
where
\begin{equation}
\widehat{T}=UTU^{T}
\end{equation}
The matrix element for matrix $\,T^{2},T^{3}\,\,and\,\,\widehat{T},\widehat{T}^{2},\widehat{T}^{3}\,$are
given in Appendix B. Using equation (48), we are in stage to define
the probability amplitude for neutrino oscillation from one flavor
to another flavor,i.e.
\[
AM_{\alpha\beta}=<\beta|e^{-iH_{f}L}|\alpha>
\]
\begin{equation}
AM_{\alpha\beta}=\varOmega\sum_{j=1}^{4}(e^{-iL\lambda_{j}}(\mathcal{L}_{1j}I_{\alpha\beta}+\mathcal{L}_{2j}\widehat{T}_{\alpha\beta}-\mathcal{L}_{3j}\widehat{T}_{\alpha\beta}^{2}-\mathcal{L}_{4j}\widehat{T}_{\alpha\beta}^{3}))
\end{equation}
Thus the transition probability is given by,
\[
P(\nu_{\alpha}\rightarrow\nu_{\beta})=AM_{\alpha\beta}^{*}AM_{\alpha\beta}=\mid AM_{\alpha\beta}\mid^{2}
\]
\[
\begin{array}{ccc}
P(\nu_{\alpha}\rightarrow\nu_{\beta}) & = & \varOmega^{*}\sum_{j=1}^{4}(e^{iL\lambda_{j}}(\mathcal{L}_{1j}I_{\alpha\beta}+\mathcal{L}_{2j}\widehat{T}_{\alpha\beta}-\mathcal{L}_{3j}\widehat{T}_{\alpha\beta}^{2}-\mathcal{L}_{4j}\widehat{T}_{\alpha\beta}^{3})^{*})\times\varOmega\\
 &  & (\sum_{k=1}^{4}(e^{-iL\lambda_{k}}(\mathcal{L}_{1k}I_{\alpha\beta}+\mathcal{L}_{2k}\widehat{T}_{\alpha\beta}-\mathcal{L}_{3k}\widehat{T}_{\alpha\beta}^{2}-\mathcal{L}_{4k}\widehat{T}_{\alpha\beta}^{3})))
\end{array}
\]
\[
P(\nu_{\alpha}\rightarrow\nu_{\beta})=\mid\sum_{j=1}^{4}(\mathcal{L}_{1j}I_{\alpha\beta}+\mathcal{L}_{2j}\widehat{T}_{\alpha\beta}-\mathcal{L}_{3j}\widehat{T}_{\alpha\beta}^{2}-\mathcal{L}_{4j}\widehat{T}_{\alpha\beta}^{3})\mid^{2}
\]

where, $\,\mathcal{L\,}$are given in eq. (38). For $\,A=A^{'}=0\,$,
we have vaccum transition probability,
\begin{equation}
P_{\nu_{\alpha}\rightarrow\nu_{\beta}}=\delta_{\alpha\beta}-4\sum_{i<j}^{4}Re(U_{\alpha i}U_{\beta j}U_{\alpha j}^{*}U_{\beta i}^{*})sin^{2}\Delta_{ij}+2\sum_{i<j}^{4}Im(U_{\alpha i}U_{\beta j}U_{\alpha j}^{*}U_{\beta i}^{*})sin2\Delta_{ij},\,\,\,\,\,\,\,\alpha,\beta=e,\mu,\tau,s
\end{equation}
Analogus to eq. (51) we can write a transition probability for (3+1)
neutrino oscillation in matter. i.e.
\[
P_{\nu_{\alpha}\rightarrow\nu_{\beta}}=\delta_{\alpha\beta}-4\sum_{i<j}^{4}Re(AM_{\alpha\beta}^{*}AM_{\alpha\beta})sin^{2}\overline{\Delta}_{ij}+2\sum_{i<j}^{4}Im(AM_{\alpha\beta}^{*}AM_{\alpha\beta})sin2\overline{\Delta}_{ij},\,\,\,\,\,\,\,\alpha,\beta=e,\mu,\tau,s
\]
\[
\begin{array}{cccc}
P_{\nu_{\alpha}\rightarrow\nu_{\beta}}= & \delta_{\alpha\beta}-4\sum_{i<j}^{4}Re(\mid\sum_{j=1}^{4}(\mathcal{L}_{1j}I_{\alpha\beta}+\mathcal{L}_{2j}\widehat{T}_{\alpha\beta}-\mathcal{L}_{3j}\widehat{T}_{\alpha\beta}^{2}-\mathcal{L}_{4j}\widehat{T}_{\alpha\beta}^{3})\mid^{2})sin^{2}\overline{\Delta}_{ij} & , & \alpha,\beta=e,\mu,\tau,s\\
 & +2\sum_{i<j}^{4}Im(\mid\sum_{j=1}^{4}(\mathcal{L}_{1j}I_{\alpha\beta}+\mathcal{L}_{2j}\widehat{T}_{\alpha\beta}-\mathcal{L}_{3j}\widehat{T}_{\alpha\beta}^{2}-\mathcal{L}_{4j}\widehat{T}_{\alpha\beta}^{3})\mid^{2})sin2\overline{\Delta}_{ij}
\end{array}
\]
where $\overline{\Delta}_{ij}=\Delta_{ij}L/2E$ , baseline length
of particular experiment is L.

\section{Conclusions}

The main results of our analysis is to determine all neutrino mass
square difference for four flavor neutrino, when neutrinos passing
through matter with constant density in Eq.(32-34). We have also calculated
the transition probabilities for neutrino oscillation in matter. We
used Cayley-Hamilton theorem be used to derive neutrino oscillation
probability. In Appendix A and B, we given coefficents and all T matrix
components,which is useful for determine neutrino oscillation probability
in matter. 

\section*{APPENDIX A : }

\subsection*{Coefficients $a_{4},\,a_{3},\,a_{2},\,a_{1}\,and\,a_{0}\,$}

\[
a_{4}=1
\]
\[
a_{3}=-Trace\,T
\]
\[
a_{3}=-(T_{11}+T_{22}+T_{33}+T_{44})=0
\]

\[
\begin{array}{cc}
a_{2}= & -[(T_{12})^{2}+(T_{13})^{2}+(T_{14})^{2}+(T_{23})^{2}+(T_{24})^{2}+(T_{34})^{2}]+T_{11}(T_{22}+T_{33}+T_{44})+T_{22}(T_{33}+T_{44})+T_{33}T_{44}\\
= & \frac{1}{8}[-3\sum_{i=2}^{4}\Delta_{i1}^{2}+2(\Delta_{21}\Delta_{31}+\Delta_{21}\Delta_{41}+\Delta_{31}\Delta_{41})+2A(\Delta_{21}(1-4U_{e2}^{2})+\Delta_{31}(1-4U_{e3}^{2})+\Delta_{41}(1-U_{e4}^{2}))\\
 & +2A^{'}(\Delta_{21}(1-4U_{s2}^{2})+\Delta_{31}(1-4U_{s3}^{2})+\Delta_{41}(1-U_{s4}^{2}))+2AA^{'}\sum_{i=1}^{4}U_{si}^{2}(1-U_{ei}^{2})-6AA^{'}\sum_{i=1}^{4}U_{ei}^{2}U_{si}^{2}\\
 & -16AA^{'}\sum_{i=1<j}^{4}U_{ei}U_{ej}U_{si}U_{sj}-3A^{2}-3A^{'2}]
\end{array}
\]

\[
\begin{array}{cc}
a_{1}= & T_{11}[(T_{23})^{2}+(T_{24})^{2}+(T_{34})^{2}-T_{22}T_{33}-T_{22}T_{44}-T_{33}T_{44}]+T_{22}[(T_{13})^{2}+(T_{14})^{2}+(T_{34})^{2}-T_{33}T_{44}]\\
 & +T_{33}[(T_{12})^{2}+(T_{14})^{2}+(T_{24})^{2}]+T_{44}[(T_{12})^{2}+(T_{13})^{2}+(T_{23})^{2}]-2(T_{12}T_{13}T_{23}+T_{12}T_{14}T_{24}+T_{13}T_{14}T_{34}+T_{23}T_{24}T_{34})
\end{array}
\]
\[
a_{0}=det(T)
\]

\[
\begin{array}{cc}
a_{0}= & (T_{12})^{2}[(T_{34})^{2}-T_{33}T_{44}]+(T_{13})^{2}[(T_{24})^{2}-T_{22}T_{44}]+(T_{14})^{2}[(T_{24})^{2}-T_{22}T_{33}]\\
 & -T_{11}[(T_{24})^{2}T_{33}+(T_{23})^{2}T_{44}+(T_{34})^{2}T_{22}-2T_{23}T_{24}T_{34}-T_{22}T_{33}T_{44}]\\
 & -2T_{12}T_{13}(T_{34}T_{24}-T_{23}T_{44})-2T_{12}T_{14}(T_{33}T_{24}-T_{23}T_{34})-2T_{13}T_{14}(T_{23}T_{24}-T_{22}T_{34})
\end{array}
\]

where, matrix elements of $\,T\,$are given in eq. (25-29).

\section*{APPENDIX B : }

\subsection*{Matrix Element for matrix $\,T^{2},T^{3}\,\,and\,\,\widehat{T},\widehat{T}^{2},\widehat{T}^{3}\,$}

\[
\begin{array}{cc}
(T^{2})_{11}= & A^{2}U_{e1}^{2}+A^{'2}U_{s1}^{2}+2AA^{'}U_{e1}U_{s1}(U_{e2}U_{s2}+U_{e3}U_{s3}+U_{e4}U_{s4})\\
 & -\frac{1}{4}(\Delta_{21}+\Delta_{31}+\Delta_{41}+A+A^{'})\left(\frac{1}{4}(\Delta_{21}+\Delta_{31}+\Delta_{41}+A+A^{'})+2(AU_{e1}^{2}+A^{'}U_{s1}^{2})\right)
\end{array}
\]
\[
\begin{array}{cc}
(T^{2})_{22}= & A^{2}U_{e2}^{2}+A^{'2}U_{s2}^{2}+2AA^{'}U_{e2}U_{s2}(U_{e1}U_{s1}+U_{e3}U_{s3}+U_{e4}U_{s4})\\
 & -\frac{1}{4}(-\Delta_{21}+\Delta_{32}+\Delta_{42})\left(\frac{1}{4}(-\Delta_{21}+\Delta_{32}+\Delta_{42}+A+A^{'})+2(AU_{e2}^{2}+A^{'}U_{s2}^{2})\right)
\end{array}
\]
\[
\begin{array}{cc}
(T^{2})_{33}= & A^{2}U_{e3}^{2}+A^{'2}U_{s3}^{2}+2AA^{'}U_{e3}U_{s3}(U_{e1}U_{s1}+U_{e2}U_{s2}+U_{e4}U_{s4})\\
 & -\frac{1}{4}(-\Delta_{32}-\Delta_{31}+\Delta_{43}+A+A^{'})\left(\frac{1}{4}(-\Delta_{32}-\Delta_{31}+\Delta_{43}+A+A^{'})+2(AU_{e3}^{2}+A^{'}U_{s3}^{2})\right)
\end{array}
\]
\[
\begin{array}{cc}
(T^{2})_{44}= & A^{2}U_{e4}^{2}+A^{'2}U_{s4}^{2}+2AA^{'}U_{e4}U_{s4}(U_{e1}U_{s1}+U_{e2}U_{s2}+U_{e3}U_{s3})\\
 & -\frac{1}{4}(-\Delta_{42}-\Delta_{43}-\Delta_{41}+A+A^{'})\left(\frac{1}{4}(-\Delta_{42}-\Delta_{43}-\Delta_{41}+A+A^{'})+2(AU_{e4}^{2}+A^{'}U_{s4}^{2})\right)
\end{array}
\]
\[
\begin{array}{cc}
(T^{2})_{12}=(T^{2})_{21}= & A^{2}U_{e1}U_{e2}\left(U_{e3}^{2}+U_{e4}^{2}\right)+A^{'2}U_{s1}U_{s2}\left(U_{s3}^{2}+U_{s4}^{2}\right)\\
 & +AA^{'}\left(U_{e3}U_{s3}+U_{e4}U_{s4}\right)\left(U_{e1}U_{s2}+U_{e2}U_{s1}\right)\\
 & +\left(AU_{e1}U_{e2}+A^{'}U_{s1}U_{s2}\right)(T_{11}+T_{22})
\end{array}
\]
\[
\begin{array}{cc}
(T^{2})_{13}=(T^{2})_{31}= & A^{2}U_{e1}U_{e3}\left(U_{e2}^{2}+U_{e4}^{2}\right)+A^{'2}U_{s1}U_{s3}\left(U_{s2}^{2}+U_{s4}^{2}\right)\\
 & +AA^{'}\left(U_{e2}U_{s2}+U_{e4}U_{s4}\right)\left(U_{e1}U_{s3}+U_{e3}U_{s1}\right)\\
 & +\left(AU_{e1}U_{e3}+A^{'}U_{s1}U_{s3}\right)(T_{11}+T_{33})
\end{array}
\]
\[
\begin{array}{cc}
(T^{2})_{14}=(T^{2})_{41}= & A^{2}U_{e1}U_{e4}\left(U_{e2}^{2}+U_{e3}^{2}\right)+A^{'2}U_{s1}U_{s4}\left(U_{s2}^{2}+U_{s3}^{2}\right)\\
 & +AA^{'}\left(U_{e2}U_{s2}+U_{e3}U_{s3}\right)\left(U_{e1}U_{s4}+U_{e4}U_{s1}\right)\\
 & +\left(AU_{e1}U_{e4}+A^{'}U_{s1}U_{s4}\right)(T_{11}+T_{44})
\end{array}
\]
\[
\begin{array}{cc}
(T^{2})_{23}=(T^{2})_{32}= & A^{2}U_{e2}U_{e3}\left(U_{e1}^{2}+U_{e4}^{2}\right)+A^{'2}U_{s2}U_{s3}\left(U_{s1}^{2}+U_{s4}^{2}\right)\\
 & +AA^{'}\left(U_{e1}U_{s1}+U_{e4}U_{s4}\right)\left(U_{e2}U_{s3}+U_{e3}U_{s2}\right)\\
 & +\left(AU_{e2}U_{e3}+A^{'}U_{s2}U_{s3}\right)(T_{22}+T_{33})
\end{array}
\]
\[
\begin{array}{cc}
(T^{2})_{24}=(T^{2})_{42}= & A^{2}U_{e2}U_{e4}\left(U_{e1}^{2}+U_{e3}^{2}\right)+A^{'2}U_{s2}U_{s4}\left(U_{s1}^{2}+U_{s3}^{2}\right)\\
 & +AA^{'}\left(U_{e1}U_{s1}+U_{e3}U_{s3}\right)\left(U_{e2}U_{s4}+U_{e4}U_{s2}\right)\\
 & +\left(AU_{e2}U_{e4}+A^{'}U_{s2}U_{s4}\right)(T_{22}+T_{44})
\end{array}
\]

\[
\begin{array}{cc}
(T^{2})_{34}=(T^{2})_{43}= & A^{2}U_{e3}U_{e4}\left(U_{e1}^{2}+U_{e3}^{2}\right)+A^{'2}U_{s3}U_{s4}\left(U_{s1}^{2}+U_{s3}^{2}\right)\\
 & +AA^{'}\left(U_{e1}U_{s1}+U_{e3}U_{s3}\right)\left(U_{e3}U_{s4}+U_{e4}U_{s3}\right)\\
 & +\left(AU_{e3}U_{e4}+A^{'}U_{s3}U_{s4}\right)(T_{33}+T_{44})
\end{array}
\]
\[
\begin{array}{cc}
(T^{3})_{11}= & AU_{e1}^{2}+A^{'}U_{s1}^{2}-\frac{1}{4}(\Delta_{21}+\Delta_{31}+\Delta_{41}+A+A^{'})((A^{2}(U_{e1}^{2}-\frac{1}{16})+A^{'2}(U_{s1}^{2}-\frac{1}{16})\\
 & -2AA^{'}(\frac{1}{16}(\Delta_{21}+\Delta_{31}+\Delta_{41}+1)-U_{e1}U_{s1}(U_{e2}U_{s2}+U_{e3}U_{s3}+U_{e4}U_{s4}))\\
 & -\frac{1}{16}(\Delta_{21}+\Delta_{31}+\Delta_{41})^{2})+(T^{2})_{12}+(T^{2})_{13}+(T^{2})_{14}
\end{array}
\]
\[
\begin{array}{cc}
(T^{3})_{22}= & AU_{e2}^{2}+A^{'}U_{s2}^{2}-\frac{1}{4}(-\Delta_{21}+\Delta_{32}+\Delta_{42}+A+A^{'})((A^{2}(U_{e2}^{2}-\frac{1}{16})+A^{'2}(U_{s2}^{2}-\frac{1}{16})\\
 & -2AA^{'}(\frac{1}{16}(-\Delta_{21}+\Delta_{32}+\Delta_{42}+1)-U_{e2}U_{s2}(U_{e1}U_{s1}+U_{e3}U_{s3}+U_{e4}U_{s4}))\\
 & -\frac{1}{16}(-\Delta_{21}+\Delta_{32}+\Delta_{42})^{2})+(T^{2})_{12}+(T^{2})_{23}+(T^{2})_{24}
\end{array}
\]
\[
\begin{array}{cc}
(T^{3})_{33}= & AU_{e1}^{2}+A^{'}U_{s3}^{2}-\frac{1}{4}(-\Delta_{32}-\Delta_{31}+\Delta_{43}+A+A^{'})((A^{2}(U_{e3}^{2}-\frac{1}{16})+A^{'2}(U_{s3}^{2}-\frac{1}{16})\\
 & -2AA^{'}(\frac{1}{16}(-\Delta_{32}-\Delta_{31}+\Delta_{43}+1)-U_{e3}U_{s3}(U_{e1}U_{s1}+U_{e2}U_{s2}+U_{e4}U_{s4}))\\
 & -\frac{1}{16}(-\Delta_{32}-\Delta_{31}+\Delta_{43})^{2})++(T^{2})_{13}+(T^{2})_{23}+(T^{2})_{34}
\end{array}
\]

\[
\begin{array}{cc}
(T^{3})_{44}= & AU_{e1}^{2}+A^{'}U_{s4}^{2}-\frac{1}{4}(-\Delta_{42}-\Delta_{43}-\Delta_{41}+A+A^{'})((A^{2}(U_{e4}^{2}-\frac{1}{16})+A^{'2}(U_{s4}^{2}-\frac{1}{16})\\
 & -2AA^{'}(\frac{1}{16}(-\Delta_{42}-\Delta_{43}-\Delta_{41}+1)-U_{e4}U_{s4}(U_{e1}U_{s1}+U_{e3}U_{s3}+U_{e2}U_{s2}))\\
 & -\frac{1}{16}(-\Delta_{42}-\Delta_{43}-\Delta_{41})^{2})+(T^{2})_{14}+(T^{2})_{34}+(T^{2})_{24}
\end{array}
\]
\[
\begin{array}{cc}
(T^{3})_{12}=(T^{3})_{21}= & AU_{e2}^{2}+A^{'}U_{s2}^{2}-\frac{1}{4}(-\Delta_{21}+\Delta_{32}+\Delta_{42})-\frac{1}{4}(A+A^{'})+(T^{2})_{12}+\left(AU_{e1}U_{e2}+A^{'}U_{s1}U_{s2}\right)(T^{2})_{11}\\
 & +\left(AU_{e2}U_{e3}+A^{'}U_{s2}U_{s3}\right)(T^{2})_{13}+\left(AU_{e2}U_{e4}+A^{'}U_{s2}U_{s4}\right)(T^{2})_{14}
\end{array}
\]
\[
\begin{array}{cc}
(T^{3})_{13}=(T^{3})_{31}= & \begin{array}{c}
AU_{e3}^{2}+A^{'}U_{s3}^{2}-\frac{1}{4}(-\Delta_{32}-\Delta_{31}+\Delta_{43})-\frac{1}{4}(A+A^{'})+(T^{2})_{13}+\left(AU_{e1}U_{e3}+A^{'}U_{s1}U_{s3}\right)(T^{2})_{11}\\
+\left(AU_{e2}U_{e3}+A^{'}U_{s2}U_{s3}\right)(T^{2})_{12}+\left(AU_{e3}U_{e4}+A^{'}U_{s3}U_{s4}\right)(T^{2})_{14}
\end{array}\\
\\
\end{array}
\]
\[
\begin{array}{cc}
(T^{3})_{14}=(T^{3})_{41}= & AU_{e4}^{2}+A^{'}U_{s4}^{2}-\frac{1}{4}(-\Delta_{42}-\Delta_{43}-\Delta_{41})-\frac{1}{4}(A+A^{'})+(T^{2})_{14}+\left(AU_{e1}U_{e4}+A^{'}U_{s1}U_{s4}\right)(T^{2})_{11}\\
 & +\left(AU_{e2}U_{e4}+A^{'}U_{s2}U_{s4}\right)(T^{2})_{12}+\left(AU_{e3}U_{e4}+A^{'}U_{s3}U_{s4}\right)(T^{2})_{13}
\end{array}
\]

\[
\begin{array}{cc}
(T^{3})_{23}=(T^{3})_{32}= & AU_{e3}^{2}+A^{'}U_{s3}^{2}-\frac{1}{4}(-\Delta_{32}-\Delta_{31}+\Delta_{43})-\frac{1}{4}(A+A^{'})+\left(AU_{e2}U_{e3}+A^{'}U_{s2}U_{s3}\right)(T^{2})_{22}\\
 & +\left(AU_{e1}U_{e3}+A^{'}U_{s1}U_{s3}\right)(T^{2})_{12}+(T^{2})_{23}+(T^{2})_{24}
\end{array}
\]
\[
\begin{array}{cc}
(T^{3})_{24}=(T^{3})_{42}= & AU_{e4}^{2}+A^{'}U_{s4}^{2}-\frac{1}{4}(-\Delta_{42}-\Delta_{43}-\Delta_{41})-\frac{1}{4}(A+A^{'})+\left(AU_{e2}U_{e4}+A^{'}U_{s2}U_{s4}\right)(T^{2})_{22}\\
 & +\left(AU_{e1}U_{e4}+A^{'}U_{s1}U_{s4}\right)(T^{2})_{12}+\left(AU_{e3}U_{e4}+A^{'}U_{s3}U_{s4}\right)(T^{2})_{23}+(T^{2})_{24}
\end{array}
\]
\[
\begin{array}{cc}
(T^{3})_{34}=(T^{3})_{43}= & AU_{e4}^{2}+A^{'}U_{s4}^{2}-\frac{1}{4}(-\Delta_{42}-\Delta_{43}-\Delta_{41})-\frac{1}{4}(A+A^{'})+\left(AU_{e3}U_{e4}+A^{'}U_{s3}U_{s4}\right)(T^{2})_{33}\\
 & +\left(AU_{e2}U_{e4}+A^{'}U_{s2}U_{s4}\right)(T^{2})_{23}+\left(AU_{e1}U_{e4}+A^{'}U_{s1}U_{s4}\right)(T^{2})_{13}+(T^{2})_{34}
\end{array}
\]
\[
\begin{array}{ccc}
\widehat{T}_{\alpha\beta}=\sum_{i=1}^{4}U_{\alpha i}\left(\sum_{j\neq i}^{4}U_{\beta j}\left(AU_{ei}U_{ej}+A^{'}U_{si}U_{sj}\right)+U_{\beta i}T_{ii}\right) & ; & \alpha,\beta=e,\mu,\tau,s\end{array}
\]
\[
\begin{array}{ccc}
\widehat{T}_{\alpha\beta}^{2}=\sum_{\alpha,\beta}(\widehat{T}_{\alpha\beta}\widehat{T}_{\beta\alpha}) & ; & \alpha,\beta=e,\mu,\tau,s\end{array}
\]
\[
\begin{array}{ccc}
\widehat{T}_{\alpha\beta}^{3}=\sum_{\alpha,\beta}\widehat{T}_{\beta\alpha}\left(\widehat{T}_{\alpha\beta}^{2}\right) & ; & \alpha,\beta=e,\mu,\tau,s\end{array}
\]

\end{document}